\documentclass[lettersize,journal]{IEEEtran}
\usepackage{amsmath,amsfonts}
\usepackage{algorithmic}
\usepackage{algorithm}

\usepackage{listings}

\usepackage{array}
\usepackage[caption=false,font=normalsize,labelfont=sf,textfont=sf]{subfig}
\usepackage{textcomp}
\usepackage{stfloats}
\usepackage{url}
\usepackage{verbatim}
\usepackage{graphicx}
\usepackage{cite}
\usepackage{bm}
\usepackage{xcolor} 
\begin{document}

\title{Calibrating DRAMPower Model for HPC: A Runtime Perspective from Real-Time Measurements}

\author{Xinyu Shi, Dina Ali Abdelhamid, Thomas Ilsche, Saeideh Alinezhad Chamazcoti, Timon Evenblij, Mohit Gupta, Francky Catthoor, \IEEEmembership{Fellow, IEEE}
 
\thanks{ Xinyu Shi, Dina Ali Abdelhamid, Saeideh Alinezhad Chamazcoti, Timon Evenblij, Mohit Gupta, and Francky Catthoor are with IMEC, Belgium. (e-mail:  shixinyu.1998@gmail.com, saeideh.alinezhadchamazcoti, timon.evenblij@imec.be, catthoor@microlab.ntua.gr, dinaabdelhamid27@gmail.com, mohitgupta.imec@gmail.com) }
\thanks{Thomas Ilsche is with TU Dresden, Germany. (e-mail: thomas.ilsche@tu-dresden.de).}
}

\maketitle
\begin{abstract}
Main memory’s rising energy consumption has emerged as a critical challenge in modern computing architectures, particularly in large-scale systems, driven by frequent access patterns, growing data volumes, and insufficient power management strategies. Accurate modeling of DRAM power consumption is essential to address this challenge and optimize energy efficiency. However, existing modeling tools often rely on vendor-provided datasheet values that are obtained under worst-case or idealized conditions. As a result, they fail to capture important system-level factors—such as temperature variations, chip aging, and workload-induced variability, which leads to significant discrepancies between estimated and actual power consumption observed in real deployments. In this work, we propose a runtime calibration methodology for the DRAMPower model using energy measurements collected from real-system experiments. By applying custom memory benchmarks on an HPC cluster and leveraging fine-grained power monitoring infrastructure, we refine key current parameters (IDD values) in the model. Our calibration reduces the average energy estimation error to less than 5\%, substantially improving modeling accuracy and making DRAMPower a more reliable tool for power-aware system design and optimization on the target server platform.

\end{abstract}

\begin{IEEEkeywords}
DRAM, HPC, power consumption, power calibration, runtime measurement.
\end{IEEEkeywords}

\vspace{-4mm}
\section{Introduction}

\IEEEPARstart{t}{he} data-intensive nature of modern applications places growing demands on computing systems, especially main memory. As computational tasks become more complex and datasets expand, DRAM not only significantly impacts system performance but also contributes heavily to total energy consumption [1]. Therefore, precise DRAM energy and power models reflecting real-world scenarios have become important for optimizing system energy efficiency. However, achieving such accuracy is challenging due to DRAM’s dynamic behavior under varying workloads and limitations of existing modeling tools.

Existing DRAM power models such as Micron’s Power Calculator [2] and DRAMPower [3], typically rely on current and voltage values provided by vendor datasheets. These parameters are measured usually under worst-case conditions on individually fabricated chips. Such datasheet-driven conditions often result in overly pessimistic estimations and fail to capture critical runtime phenomena, including chip aging, temperature changes, and other sources of variability, leading to substantial errors when estimating real-system energy consumption.

Previous calibration method Vampire [4], has experimentally characterized DDR3-standard DRAM current on FPGA-based memory-controller simulators using off-the-shelf memory modules. More recent research [5] has explored architectural innovations to improve DRAM energy efficiency but lacks the detailed calibration of DRAM current parameters offered by Vampire. However, existing FPGA-based calibration methods typically provide fine-grained control enabling precise calibration under tightly controlled conditions, which is unsuitable for large-scale HPC clusters. In contrast, HPC servers operate under kernel-driven memory management, restricting direct user control over memory allocation and access patterns, and involve adaptive refresh strategies and complex temperature variations. Therefore, directly applying FPGA-derived IDD values to server scenarios risks substantial inaccuracies in power estimation. To the best of our knowledge, no existing work has addressed these HPC-oriented calibration challenges, motivating the need for a complementary calibration approach explicitly designed for realistic HPC operating environments.

To bridge these gaps, we propose a calibration methodology that leverages power measurements from a real HPC server equipped with the High Definition Energy Efficiency Monitoring (HDEEM) system [6], combined with a full-system-level memory simulator, to calibrate the supply currents (IDD) in the DRAMPower model. Our contributions include: (1) Using the existing Taurus HPC infrastructure at TU Dresden and custom-designed memory benchmarks, we systematically analyze DRAM power characteristics under various parallel workloads, revealing how parallelism and memory allocation impact DRAM energy consumption; and (2) we propose and demonstrate a DRAMPower IDD calibration method that can be directly deployed on HPC clusters without additional hardware, validated by a detailed calibration case study using actual server measurements, significantly improving model accuracy. The rest of the paper is organized as follows: Section II explains the methodology and experimental setup; Section III analyzes the runtime power measurement results; in Section IV we present a DRAMPower calibration case study and insight; and Section V discusses limitations, future work and conclusion of this work.
\vspace{-2mm} 
\section{Methodology and Experimental Setup}
This session outlines the methodology behind the work, including a brief introduction to the HPC infrastructure, the benchmark, and the simulators that enable the calibration. Fig. 1 provides an overview of the workflow employed in this study.
\vspace{-2mm}
\begin{figure*}[htbp]
\centering
\includegraphics[width=0.95\textwidth]{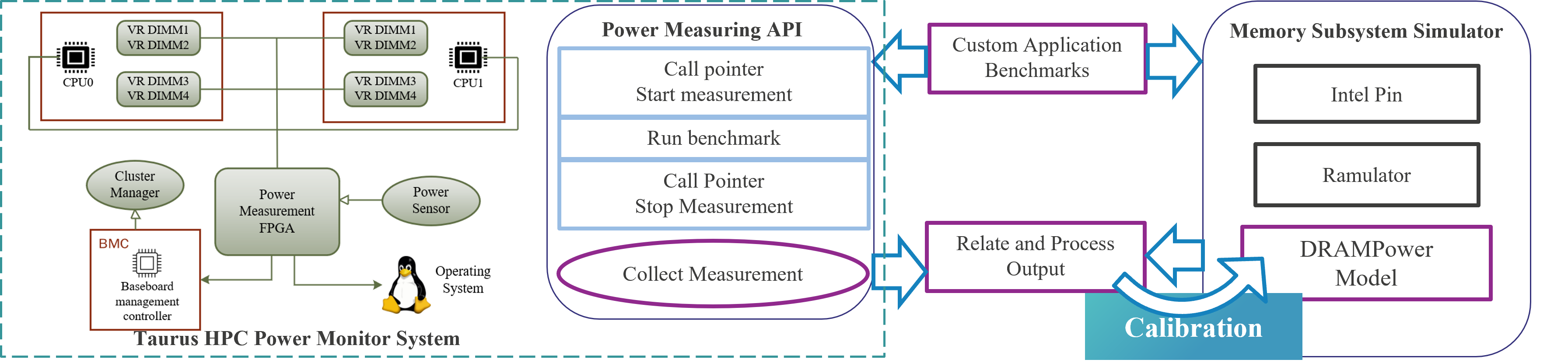} 
\vspace{-3mm} 
\caption{Overall workflow of HPC power measurement system and DRAMPower calibration. The left side illustrates the HPC measurement system (including HW configuration and API), while the right side presents the configuration of the simulators.}
\label{fig_1}
\end{figure*}
\vspace{-3mm}

\subsection{Server Infrastructure}
HDEEM is an integrated hardware-software solution that accurately monitors real-time power and energy consumption throughout the system using APIs that interface with power sensors around the CPU and memory, with data processing handled by an FPGA. Our experiments are deployed on compute nodes in the Haswell partition of the Taurus HPC system. To ensure that power measurements are not contaminated by background activity, all experiments are submitted via SLURM in exclusive mode, reserving full compute nodes during execution. Each compute node is configured with dual CPUs that has maximum 24 cores running in parallel, 32MB of L3 cache and 64GB of DDR4 main memory, using SAMSUNG M393A1G43DB0-CPB modules.\footnote{Some nodes might have different DIMM modules. Here we only look at the node with this specific configuration.} Each memory DIMM has an 8GB capacity, organized with 2 ranks and 16 banks. The 8 DIMMs are grouped into four pairs, each pair being monitored independently by dedicated power sensors.
\subsection{Benchmarks}
B. Benchmark
Both measurement and calibration require extensive memory access to maximize utilization, while maintaining a simple and stable access pattern for easy analysis and calibration; we customize benchmarks based on the STREAM memory bandwidth benchmark [7]. STREAM is specifically designed for memory-intensive workloads, which helps to generate high- memory access traffic. Meanwhile, it provides simple, predictable memory access patterns that make it easier to analyze and correlate power consumption with specific memory behaviors. Table 1 lists some examples of the benchmarks we use and their access patterns used in this work.
\begin{table}[t]
  \caption{Benchmark Access Pattern}
  \label{tab:bench-access}
  \centering
  \setlength{\tabcolsep}{8pt}
  \renewcommand{\arraystretch}{1.2}
  \begin{tabular}{|l|l|}
    \hline
    \textbf{Benchmark} & \textbf{Access Pattern} \\
    \hline
    Read      & $sum += a[j]$ \\
    Assign    & $a[j] = 2$ \\
    Scale     & $b[j] = 2 \times a[j]$ \\
    Addition  & $c[j] = a[j] + b[j]$ \\
    Triad     & $a[j] = b[j] + 2 \times a[j]$ \\
    Copy      & $b[j] = a[j]$ \\
    SelfScale & $a[j] = 2 \times a[j]$ \\
    \hline
  \end{tabular}
\end{table}

\lstdefinestyle{membench}{
  language=C,
  basicstyle=\ttfamily\footnotesize,  
  numbers=left, numberstyle=\tiny, stepnumber=1, numbersep=4pt,
  frame=single, rulecolor=\color{black},
  columns=fixed, keepspaces=true,
  breaklines=true, breakindent=0pt, breakatwhitespace=false,
  literate        = {_}{{\_}}1
                    {.}{{.}}1,
  captionpos=b,                       
  postbreak=\mbox{$\hookrightarrow$}, 
}

\begin{lstlisting}[style=membench,
    caption={\textbf{Read} kernel (stride = 8) used in our STREAM-like benchmark},
    label={lst:read_kernel}]
#define ARRAY_SIZE 100000000
...
void read_kernel(double *a, double sum) {
    sum = 0.0;
    for (long k = 0; k < ARRAY_SIZE; k += 8)
         sum += a[k];         
}
\end{lstlisting}

\vspace{-3mm}
\subsection{Simulator and Framework}
To accurately calibrate the DRAMPower model, we select tools that enable the accurate mimicry of the real system's dataflow and memory access. We select Intel Pin Tool[8] to generate CPU traces from the benchmark on the real system, and Ramulator[9] to simulate the functional behavior of the main memory. In our simulation, these tools use simplified representations of system-level pipelining and concurrency, which may not fully reflect the complex access patterns and mappings of real HPC systems.
\subsubsection{Intel Pin}
Intel Pin [8] is a system-level dynamic binary instrumentation tool that analyzes program behavior at runtime by inserting custom routines into executable without modifying the source code. In this work, Intel Pin is used on the Taurus HPC cluster to generate CPU traces, providing detailed information on executed instructions and memory read/write operations.
\subsubsection{Ramulator}
Ramulator [9] is a cycle-accurate DRAM simulator that translates the CPU traces generated from instrumentation into memory-controller-level commands based on the DDR standard, memory specification parameters and the system’s mapping rules. This outputs memory-operation-level command traces.
\vspace{-3mm}
\subsection{DRAMPower Model}
DRAMPower[3] is a tool for modeling DRAM energy and power consumption based on application-driven memory command traces. It processes command traces produced by simulators like Ramulator and performs statistical analysis by counting various types of memory operations in the trace. Using detailed timing specifications from memory configuration files, DRAMPower provides an accurate assessment of power and energy consumption for different DRAM operations. 

The behavior of DRAM and the execution of its commands are driven by different IDD currents. Each memory operation is associated with one or more IDD currents, which are calibrated in our work. A single memory command’s simulated energy in DRAMPower Model can be simplified as equation(1), where T\textsubscript{op} stands for related timing spec of the command; N\textsubscript{bank} stands for the number of banks in the DIMM, I\textsubscript{op} stands for the currents related to the command, and N\textsubscript{cmd} is the total numbers of this particular command.
\begin{equation}
\label{deqn1}
 E_\text{cmd} = V_{\text{dd}} \times T_\text{op} \times N_\text{bank} \times I_\text{op} \times N_\text{cmd}
\end{equation}
The overall average power consumption of the application is formalized as equation(2), which stands for the total simulated energy of the application divided by the total simulated time, where C\textsubscript{total} is the total numbers of clock cycles, and t\textsubscript{Clk} is the system clock.
\vspace{-3mm}
\begin{equation}
    P_\text{avg} = \frac{\sum_{0}^{n} E_\text{cmd}}{C_\text{total} \times t_\text{Clk}}
\end{equation}
\vspace{-5mm}
\section{Runtime Power Measurement}
In this section, we analyze runtime power measurements based on our custom benchmarks. As a prerequisite for effective calibration, the memory workload executed on the real server must mirror the one used in the simulator so that both runs exercise the same access pattern. Because users on our HPC system cannot reconfigure the OS allocator or the memory-controller interleaving policy, we reverse-engineer the address mapping by analyzing the physical-address traces (from Pin) so that this can be used to configure the simulator with the inferred channel/rank/bank/row/column mapping.

Meanwhile, standalone DRAM-level simulators do not model CPU caches or TLB/page-table walks. To minimize interference from such system effects and ensure consistent memory behavior for calibration, we allocate arrays much larger than the last-level cache so most accesses reach DRAM, and we use a minimal C loop structure while preventing compiler optimizations from eliminating the memory operations. We fix the loop stride to 64 B (8 doubles) so that each iteration touches exactly one cache line (one DRAM burst with BL=8), keeping TLB effects negligible under 4 KB pages and yielding a deterministic, cache-agnostic DRAM request stream.
\vspace{-3mm}
\subsection{Active Power and Parallelization impact}
To study active power consumption of given memory workload and how parallelism affect DRAM's behaviour on HPC server, we measure the average DIMM power while gradually increasing the number of active threads from 1 to maximum 24. Each thread runs an identical micro-benchmark, and the resulting power-versus-thread curve allows us to know the operating system’s memory-allocation policy rather than setting it explicitly. Fig. 2 plots the four measurement channels (DDR AB, CD on socket 0 and DDR EF, GH on socket 1) for the Addition benchmark. \footnote{Due to space limitations, here we only demonstrate one benchmark's result. Although other benchmarks exhibit different power consumption results due to varying execution patterns, the overall trend of power consumption increases with the number of parallel threads remains similar.}

\begin{figure}[htbp]
\centering
\includegraphics[width=\columnwidth]{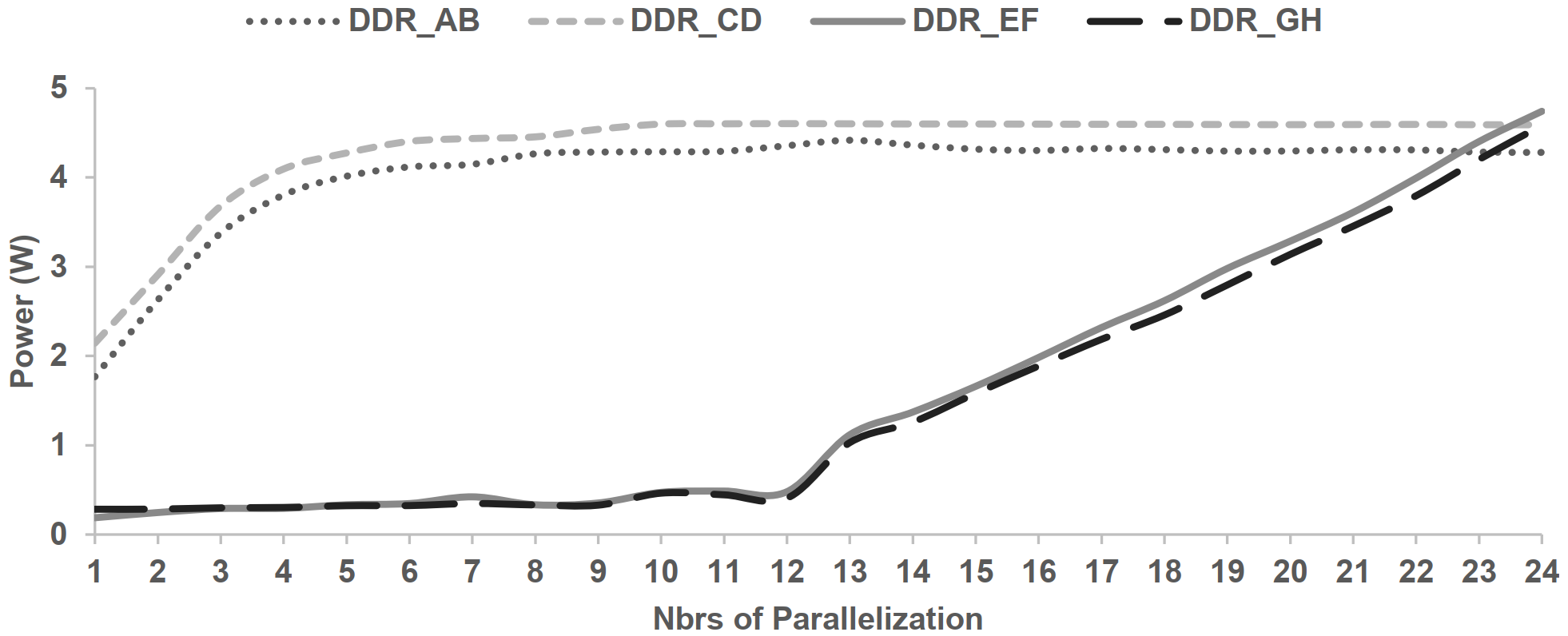}
\vspace{-4mm}
\caption{Average Power Consumption of Addition Benchmark.}
\label{fig_2}
\end{figure}

\vspace{-1mm}
As shown in Fig. 2, DRAM channels DDR AB and CD exhibit a nearly linear increase in power consumption from one to approximately four parallel threads, followed by a graceful slow down till eight threads. This behavior reflects a proportional rise in memory accesses until the bandwidth of memory channels controlled by socket 0 becomes gradually saturated. When thread count exceeds 12, the variables declared in the thread code cannot be allocated to socket 0 any more by the linker due to capacity limitations. So additional threads are assigned onto the second CPU socket, activating DRAM channels EF and GH. Since the initial 12 threads still continue generating memory traffic on channels AB/CD, the two sockets concurrently handle the workload. As a result, channels EF and GH approach their bandwidth-limited power plateau only near the maximum tested parallelism of 24 threads, highlighting the staged utilization and saturation pattern across the two sockets. The differences in slopes and absolute power values for the different channels are attributed to variability (see earlier).

\subsection{Static Power}
Apart from understanding dynamic power during active workload execution, it's also important to characterize the DRAM’s static power consumption when no active memory operations are being performed. This helps us identify baseline idle power levels and isolate the underlying leakage and background maintenance activities of the memory system. To minimize memory activity while keeping the CPU fully active, we design a compute-only benchmark using local register-based operations, which avoids global memory access and ensures that DRAM remains in a standby-refresh state during execution.

Fig. 3 illustrates the measured static power consumption across four monitoring channels as an instantaneous-power-versus-pulling-duration curve. As can be observed from Fig. 3, each channel’s power readings exhibit slight fluctuations over time and a stable per-channel offset. The short-term fluctuations are mainly due to DIMM temperature drift and supply-rail noise (VDD ripple), whereas the persistent offset between channels (e.g., AB vs. CD/EF/GH) is consistent with non-uniform cooling and board-level placement reported for this server. Accordingly, we adopt a server-specific, per-channel baseline: over the measurement window, the average static current is approximately 0.166 A per DIMM, which provides a baseline for leakage-dominated current draw in idle DRAM states within the studied server environment.

\begin{figure}[htbp]
\centering
\includegraphics[width=\columnwidth]{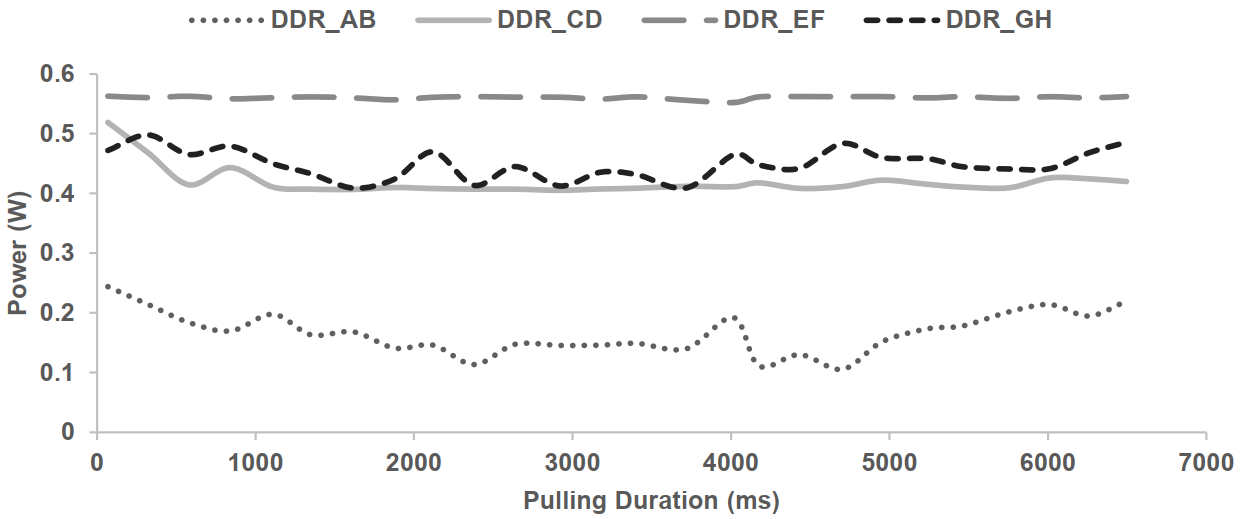} 
\vspace{-4mm} 
\caption{Runtime static power consumption over polling duration.}
\label{fig_3}
\end{figure}
\vspace{-5mm}
\section{DRAMPOWER Model Calibration}
Building on the runtime measurements in Section III, in this section, we present our calibration methodology by showing a case study based on our benchmarks and memory module mentioned previously to calibrate the DRAMPower supply current parameters (IDDs) using a regression-based approach. As described in Section II-C, the energy of each DRAM command can be represented as a sum of the product of various current parameters and timing factors. Summing across all commands in a given benchmark, the total energy can be reformulated as a linear combination of the IDDs plus a small intercept that absorbs baseline bias as equation 3, \begin{equation}
\label{eq:linear-idd}
E_{\text{total}}^{(k)} = \sum_{i} \mathrm{Coeff}_i^{(k)} \cdot \mathrm{IDD}_i \; (+\, b),
\end{equation}
where $\mathrm{Coeff}_i^{(k)}$ aggregates all non-current factors (timings, rank/bank counts, and command statistics), and b absorbs residual baseline bias such as the I/O energy which are not determined by IDD currents. Using measured energy under 24-thread parallel execution as the ground truth, we solve for the IDDs that best fit the observations under sustained high memory load.

\begin{figure*}[htbp]
  \centering
  \subfloat[\scriptsize]{\includegraphics[width=0.48\linewidth]{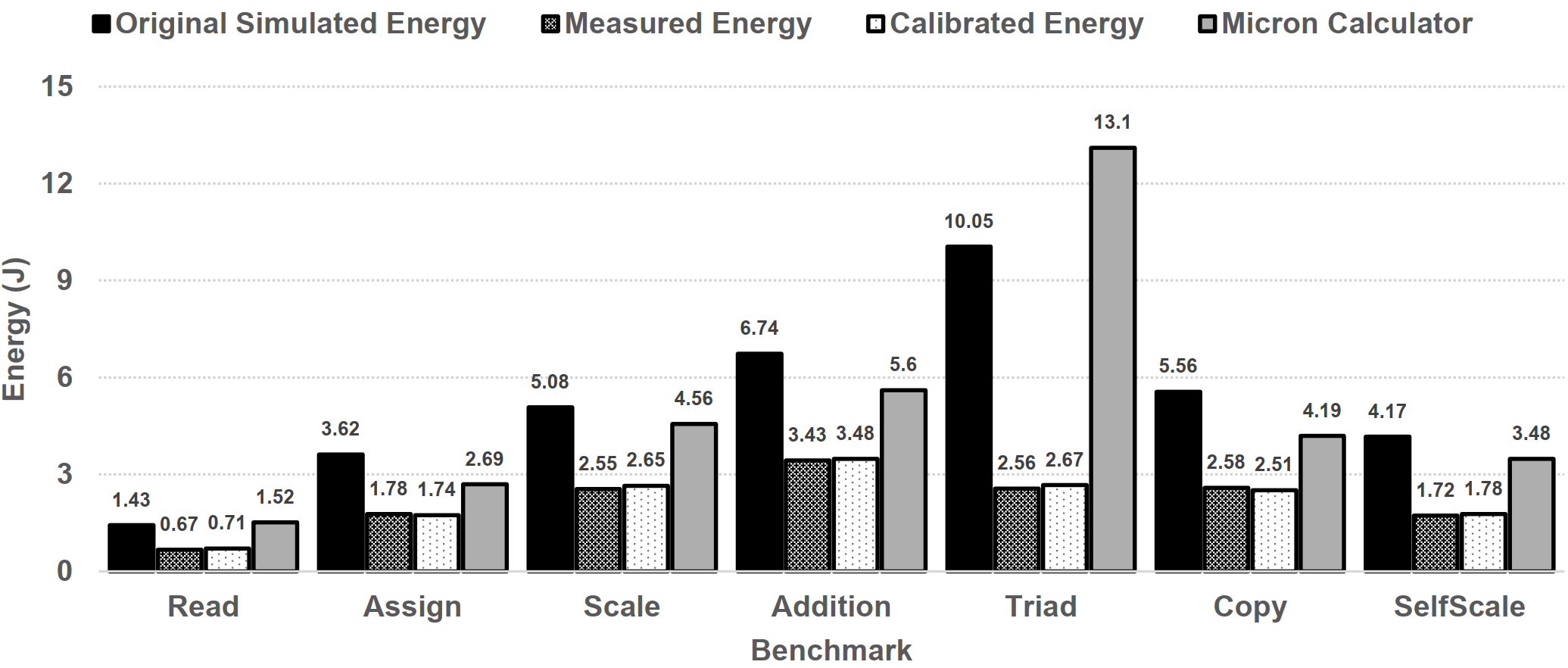}}
  \hfill
  \subfloat[\scriptsize]{\includegraphics[width=0.48\linewidth]{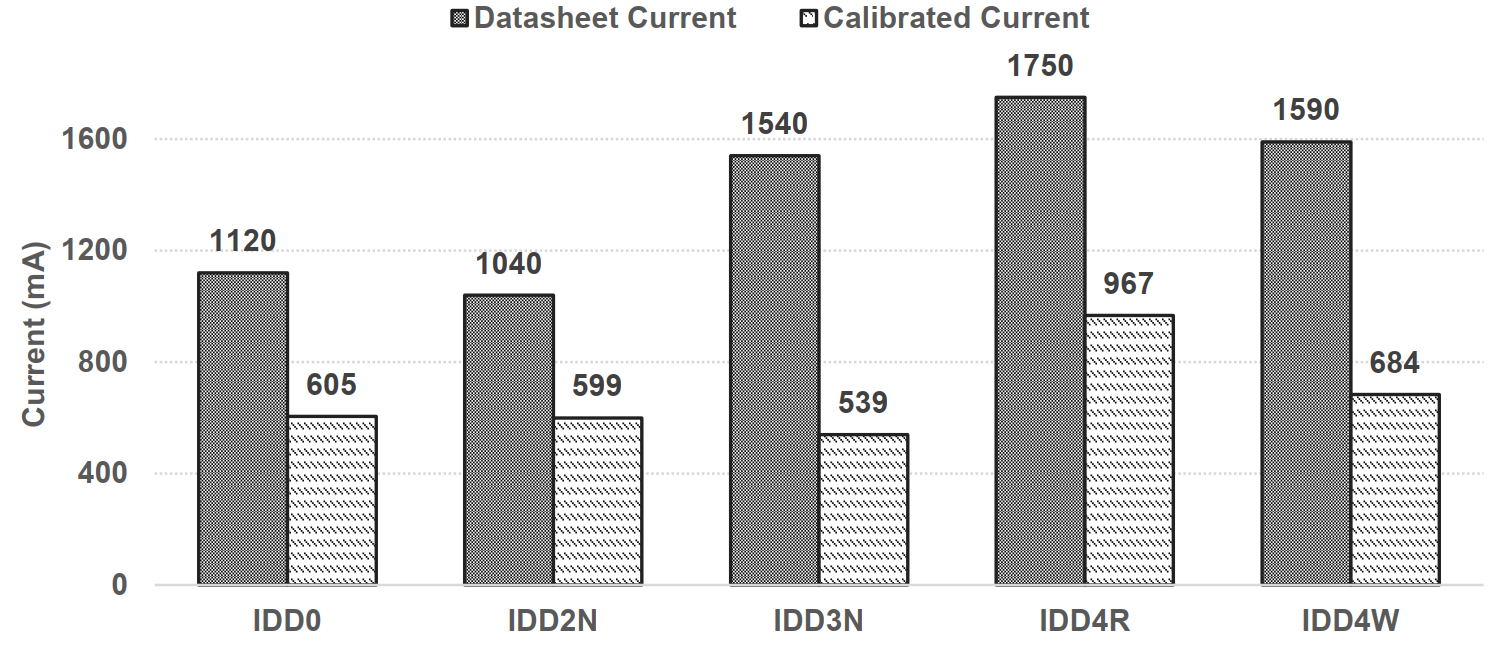}}
  \caption{(a) Comparison between measured, simulated, calibrated, and Micron energy values. (b) Original datasheet IDD currents vs. calibrated values.}
  \label{fig:calibration}
\end{figure*}
Figure 4(a) presents an aggregate comparison between simulated (uncalibrated and calibrated) and measured energy. The uncalibrated DRAMPower model consistently overestimates energy for all benchmarks; after regression-based calibration of five key currents (activation, pre-charge, active-standby, read, and write), Figure 4(b) shows the updated values. Applying these currents to DRAMPower yields post-calibration simulations that closely align with measurements, reducing the average error to less than 5\%. For reference, we also show results from the vendor power calculator, using the same speed bin and rank configuration.

\lstdefinestyle{lsq}{
  language=Python,
  basicstyle=\ttfamily\footnotesize,
  numbers=left, numberstyle=\tiny, stepnumber=1, numbersep=4pt,
  frame=single, rulecolor=\color{black},
  keepspaces=true, columns=fixed,
  breaklines=true, breakatwhitespace=true,
  literate        = {_}{{\_}}1
                    {.}{{.}}1,
  captionpos=b,                       
  postbreak=\mbox{$\hookrightarrow$},
}

\begin{lstlisting}[style=lsq]
import numpy as np
from scipy.optimize import lsq_linear

# Build A and b
A = VDD * N * T        # shape: (K, 5)
b = E_meas - E_const   # shape: (K,)

# Bounds 0 <= I <= IDD_max
lower = np.zeros(5)
upper = IDD_max

# Solve
I_star = lsq_linear(A, b, bounds=(lower, upper), method='trf').x
\end{lstlisting}

The calibration problem is formulated as a bounded least-squares optimization: stacking each benchmark’s coefficient vector as a row in matrix $\mathbf{A}$ and measured energy as a vector $\mathbf{y}$, we minimize
\begin{equation}
\min_{\boldsymbol{\mathrm{IDD}}}\ \tfrac{1}{2}\,\big\lVert \mathbf{A}\,\boldsymbol{\mathrm{IDD}} - \mathbf{y} \big\rVert_2^2
\quad \text{s.t.}\quad \boldsymbol{\ell} \le \boldsymbol{\mathrm{IDD}} \le \boldsymbol{u}.
\label{eq:bounded-lsq}
\end{equation}
subject to physical bounds on each current (reflecting datasheet values).
The pesudo code listed here shows the implementation of our algorithm. For each benchmark, we calculate the appropriate coefficients from access statistics and timing to build the matrix $\mathbf{A}$ (with columns corresponding to the five currents). Terms independent of IDDs are grouped as constants and subtracted from measured energy to form the right-hand side $\mathbf{y}$. We set lower and upper bounds for each current (e.g., zero to the datasheet maximum) and use a bounded least-squares solver such as \texttt{lsq\_linear} with the \texttt{trf} algorithm, which ensures that solutions are both physically meaningful and numerically stable---even when some columns in $\mathbf{A}$ are highly correlated.

\begin{figure}[htbp]
\centering
\includegraphics[width=\columnwidth]{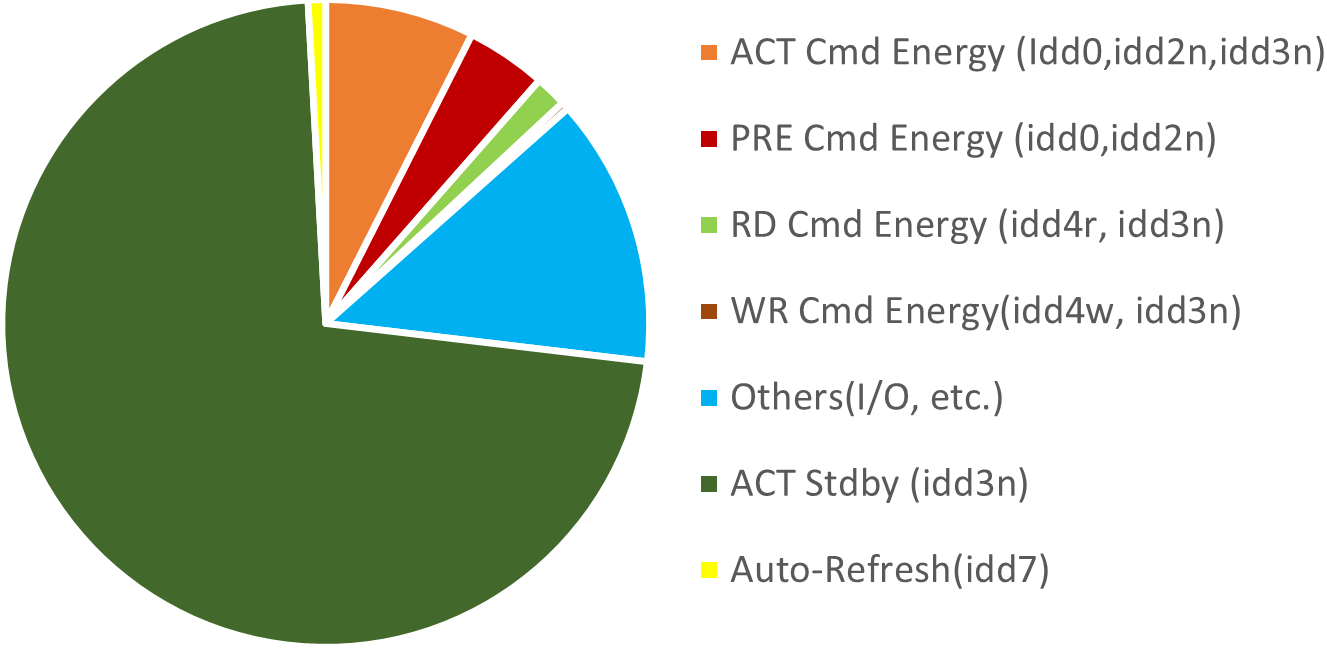}
\vspace{-4mm}
\caption{Add Benchmark Original Simulated Energy Breakdown.}
\label{fig_2}
\end{figure}
As we can see from the figure 4.(b), only several Idd currents here are calibrated. The pie chart Fig. 5 shows the add benchmark's simulated energy breakdown from the original DRAMPower model, with different memory commands driven by multiple Idd currents, which highlights issues of calibration observability rather than merely energy composition. 
DRAMPower decomposes energy into two parts: (i) background energy due to time spent in baseline states (IDD2N/IDD3N), 
and (ii) incremental energy from commands above that baseline (e.g., for read/write, the delta over IDD3N). 
For memory-intensive workloads, the dominant error contributions come from the incremental activation and read/write currents, 
while several currents (e.g., refresh, I/O, rare modes) make up only a small fraction of the total energy. 
These ``small-slice'' parameters are difficult to calibrate for several reasons:
\begin{itemize}
    \item \textbf{Weak excitation / low signal-to-noise:} Their coefficients in $A$ are small and easily masked by measurement noise, providing little leverage in regression.
    \item \textbf{Col-linearity:} Their columns in $A$ are nearly linear combinations of dominant ones (e.g., dwell-time terms), making solutions non-unique unless bounded.
    \item \textbf{Duty-cycle constraints:} Some operations occur so infrequently at full load that, unless specifically targeted by microbenchmarks or temperature sweeps, the available data are insufficient to independently determine them.
\end{itemize}
Consequently, the bounded solver keeps these weakly observable currents close to datasheet values, while the main corrections are concentrated on the strongly excited parameters---namely, activation, read/write, and active-standby. 
The purpose of the pie chart is to illustrate this point: small-slice currents are not ``unimportant,'' but are difficult to reliably calibrate given the available measurement data.

\begin{figure}[htbp]
\centering
\includegraphics[width=\columnwidth]{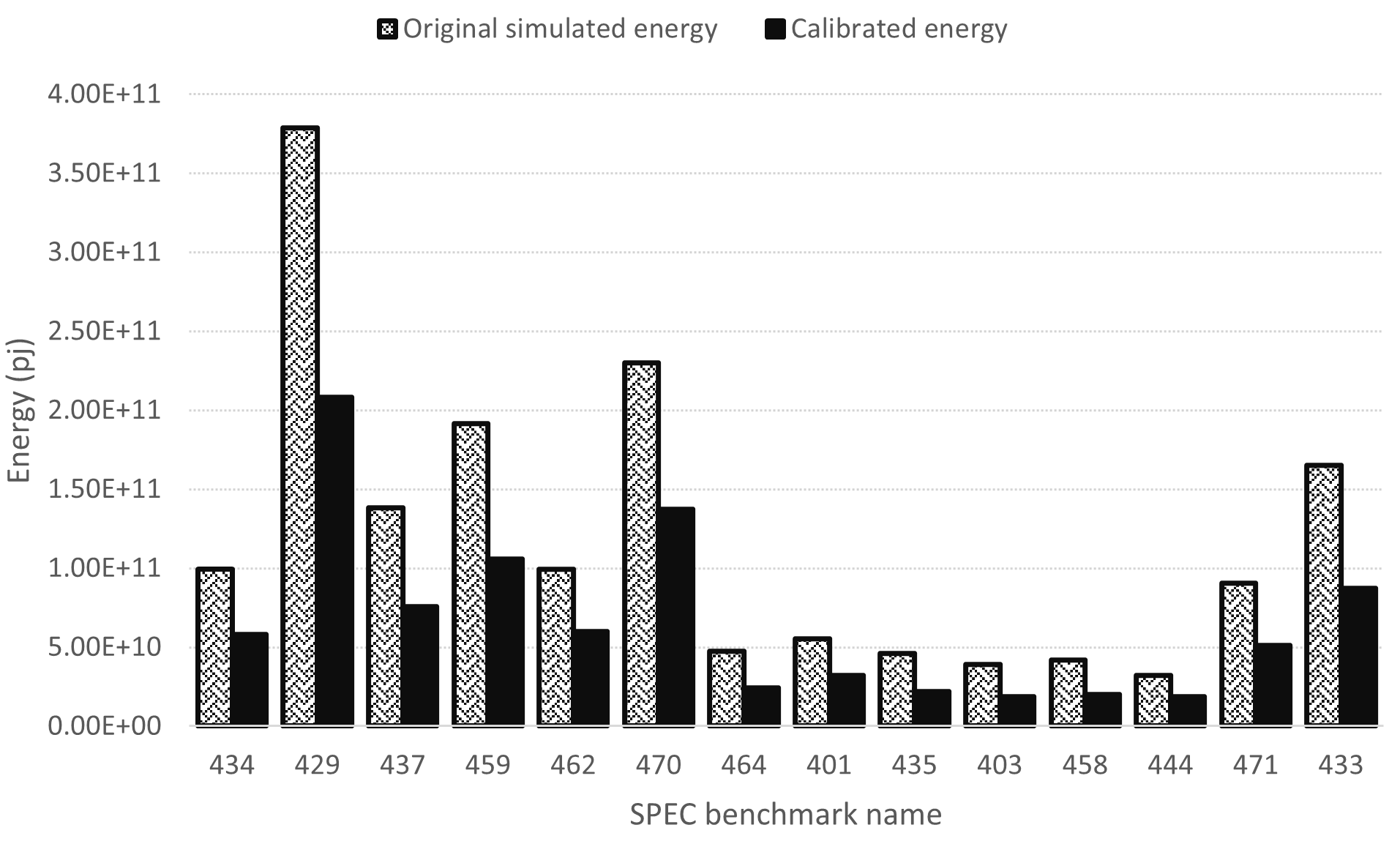}

\caption{SPEC2006 Original and Calibrated Energy.}
\label{fig_5}
\end{figure}
 To evaluate generalization, we apply the calibrated currents to a set of SPEC2006 workloads. Figure 6 compares simulated energy before and after calibration. Most benchmarks show simulated energy shifting toward the measured scale, typically reducing pre-calibration estimates by roughly half—consistent with the bias correction seen in Figure 4(a). This confirms that, without altering the underlying model structure, a single calibration pass can deliver stable accuracy gains across diverse workloads and load intensities.

\section{Discussion and Conclusion}
In this paper, we presented a methodology for calibrating the DRAMPower model's IDD currents using power measurements from an HPC environment, demonstrating its practical applicability and effectiveness. However, only five IDD currents listed in Fig. 4(b) are successfully calibrated. While additional IDD parameters influence DRAM behaviors, some cannot be accurately calibrated using our approach due to two main limitations. First, measurements obtained from HPC environment inherently contain non-negligible noise as shown in Fig. 3. Although averaging multiple runs helps mitigate this, residual measurement noise remains inevitable. Second, the DRAMPower tool primarily simulates active memory operations based on memory traces. Consequently, currents associated with less frequent operations, such as refresh events, have limited data for accurate calibration and therefore remain uncalibrated.

Meanwhile, compared to previous FPGA-based approaches such as VAMPIRE, which provide fine-grained control over DRAM operations and precise environmental conditions (e.g., controlled temperature sweeps), our methodology offers a complementary calibration specifically suited for realistic, long-running HPC scenarios. As detailed earlier, the controlled FPGA environments fundamentally differ from HPC servers, where kernel-managed memory allocations, adaptive refresh strategies, and complex thermal conditions prevail. 
Our method uses existing HPC infrastructure, enabling practical, cost-effective calibration for systems that have operated extensively in real-world conditions.
Consequently, FPGA-based methods and our HPC-based calibration mutually complement each other, together delivering a comprehensive and versatile strategy for accurate DRAM power modeling across a variety of computing platforms.
\vspace{-2mm}

\section{Acknowledgments}
The authors gratefully acknowledge the computing time made available to them on the high-performance computer at the NHR Center of TU Dresden. This center is jointly supported by the Federal Ministry of Education and Research and the state governments participating in the NHR(www.nhr-verein.de/unsere-partner).

\vfill

\end{document}